# ROBOTIKANDO: a Web Tool for Supporting Teachers Practicing Robotics in Kindergarten

**Ornella Mich**

Fondazione Bruno Kessler
Trento, 38123, I
mich@fbk.eu

**Roberto Tiella**

Fondazione Bruno Kessler
Trento, 38123, I
tiella@fbk.eu

## Abstract
This paper describes ROBOTIKANDO, a web application for supporting both kindergarten teachers in planning educational robotics activities and educational robotics experts who would like to share their knowledge and experience. ROBOTIKANDO has been designed and implemented following a co-design process, which devised a conceptual map aiming to connect educational robotics and kindergarten education principles. As future work, we are planning a longitudinal evaluation with Preschools teachers. Moreover, we are thinking of extending the application to teachers of primary and secondary schools.

## ACM Classification Keywords
H.5.m. Information interfaces and presentation (e.g., HCI): Miscellaneous.

## Introduction
Educational robotics is the set of all those educational activities based on the design, creation, assembly and operation of small robots. Educational robotics is beneficial to students starting from primary school up to high school. Major educational theories [4, 5] lie at the basis of educational robotics. Several studies [1, 3, 6] have demonstrated that educational robotics supports and reinforces specific areas of knowledge (Physics, Mathematics, Engineering and Informatics) and cognitive, meta-cognitive and social skills (research skills, creative thinking, decision making, problem solving, communication and team working skills). Educational robotics also has the potential for eliminating barriers to learning in children with special needs [8]. Recently, attempts to introduce robotics at kindergarten have also been documented [2, 3, 7]. However, kindergarten teachers (at least in Italy), often missing a background in educational robotics, can be initially puzzled as to how employ robotics as a learning tool. This paper describes ROBOTIKANDO, a web tool for supporting kindergarten teachers in introducing robotics practices in their classrooms. ROBOTIKANDO is an activity planner [9], specifically designed for teachers inexperienced in robotics, who would like to start introducing robotics in their classrooms in a systematic way. But our application is also designed for educational robotics experts who would like to share their knowledge and experience with colleagues to have feedback and suggestions to improve their research. ROBOTIKANDO has been designed and implemented following a co-design process (see Figure 1).

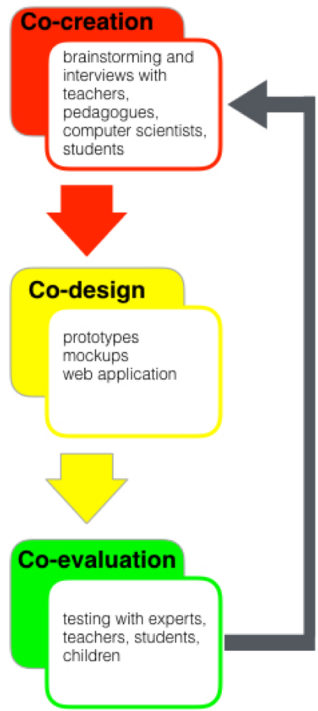

**Figure 1**: The design process of ROBOTIKANDO.

## Application description

Figure 2 presents the Home page of a prototype of ROBOTIKANDO. Its key features are represented in Table 1. ROBOTIKANDO was designed starting from the conceptual map shown in Figure 3. This map represents the connection between the main concepts that are the core of our vision on how to introduce educational robotics in kindergarten. The leading concept for an Italian kindergarten teacher is to address an *educational topic* a year, for example *Collaboration* or *Inclusion*, etc. Using our application, teachers wanting for example to convey the importance of collaboration to their students, will organize activities that will place children in collaborative situations. To this aim, the teacher using our application will choose a *Scenario*, namely an environment, as abstract as mathematics or as physical as an amusement park for example, where the activities will take place. Activities are full lesson plans proposed to and realized by children, such as building a robotic merry-go-round in the amusement park scenario. Activities require robotic and programming skills (What) such as understanding how sensors and motors work and how they can be used in programs. Activities can be distributed to groups of different sizes, ranging from children working on their own to whole-class experiences. Finally, "what" aspects, such as motor and sensors, will be embodied in specific tools and kits. It is worth noting that we also added a "WHO" perspective to address, in future implementations, students of different ages. Following the application work flow, teachers are guided to collect the more suitable material and information to build effective lessons.

## Future work

As future work, we are planning to move from the prototype to a completely functioning application. We are also planning a longitudinal evaluation for measuring the perceived usefulness of the application with teachers working in the 133 Preschools associated to the Provincial Federation of Preschools of Trento, Italy. With these teachers, we will also investigate whether it is correct to propose the same activity to a group of mixed aged children (from 3 to 5), as currently implemented, or whether children's age should be considered in planning the activities. Moreover, we are thinking of extending the application to teachers of primary and secondary schools, as already reflected in the conceptual map.

| ROBOTIKANDO features |
| --- |
| **driven by the conceptual map** |
| **personalized content**: the application keeps track of activities the teacher had proposed and suggests next steps |
| **proactiveness:** the application provides suggestions that are related to past activities by means of relationships in the conceptual map |
| **collaborative environment**: the application allows users to add comments and suggestions to activities in order to share their experience. In this way, they can help and advice colleagues and improve the quality of application's knowledge base. |

**Table 1**. The main features of ROBOTIKANDO.

## Acknowledge

We thank Nicola De Bortoli and Francesco Piccolboni for creating the ROBOTIKANDO prototype. We also thank Camilla Monaco and Tiziana Ceol, researchers at the Provincial Federation of Preschool of Trento, Italy, for their support in the definition of the key features of ROBOTIKANDO.

## References

1. Barker, B.S., & Ansorge, J. Robotics as means to increase achievement scores in an informal learningenvironment. J. of Research on Technology in Education, 39(3), (2007), 229-243.
2. Marina Umaschi Bers, Flannery, L., Kazakoff, E.R. and Amanda Sullivan. 2014. Computational thinking and tinkering: Exploration of an early childhood robotics curriculum. *Computers & Education*, 72, 145-157.
3. Druin, A., & Hendler, J. A. Robots for Kids: Exploring New Technologies for Learning. (Eds). Morgan Kaufmann. (2000).
4. Papert, S. Mindstorms: Computers, Children and Powerful Ideas. NY: Basic Books, (1980).
5. Piaget, J. *The Child's Conception of the World*. (1929).
6. Resnick, M. Technologies for lifelong kindergarten. *Educational Technology Research and Development*, 46(4), (1998), 43-55.
7. Stoeckelmayr, K., Tesar, M., & Hofmann, A. Kindergarten children programming robots: a first attempt. *Proc. of 2nd Int. Conf. on Robotics in Education* (RiE 2011) (2011), 185-192
8. Virnes, M., Sutinen, E., & Kärnä-Lin, E. How children's individual needs challenge the design of educational robotics. *Proc. of the 7th Int, Conf. on Interaction Design and Children*. ACM. (2008, June), 274-281.
9. Yiannoutsou, N., Nikitopoulou, S., Kynigos, C., Gueorguiev, I., & Fernandez, J. A. (2017). Activity plan template: a mediating tool for supporting learning design with robotics. In *Robotics in Education* (pp. 3-13). Springer, Cham.

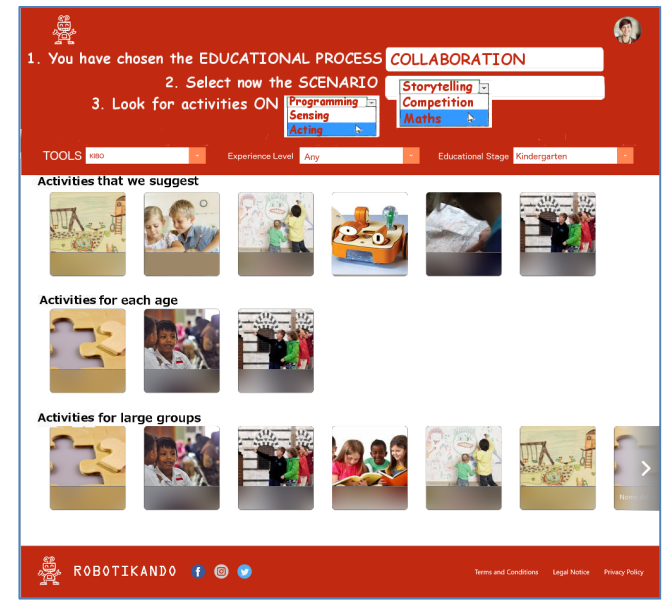

**Figure 2**: The Home page of ROBOTIKANDO.

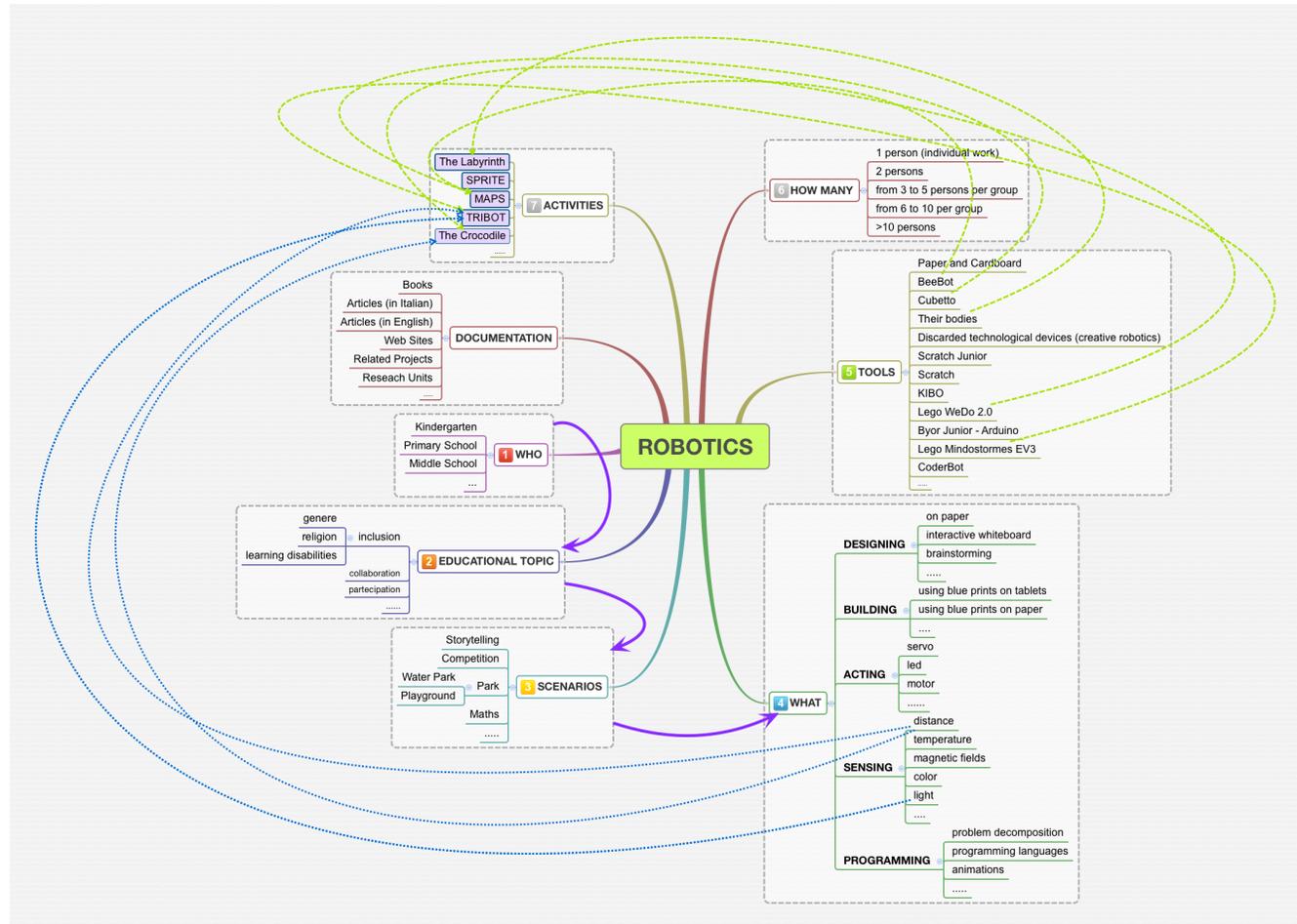

**Figure 3**. Conceptual Map representing the structure of ROBOTIKANDO (simplified – not all the connections are drawn for the sake of readability).